\documentstyle [12pt] {article}
\input epsf

\newcommand{\gtwid}{\mathrel{\raise.3ex\hbox{$>$\kern-.75em\lower1ex
\hbox{$\sim$}}}}
\newcommand{\ltwid}{\mathrel{\raise.3ex\hbox{$<$\kern-.75em\lower1ex
\hbox{$\sim$}}}}
\newcommand{\beq}{\begin{equation}}
\newcommand{\eeq}{\end{equation}}
\newcommand{\beqs}{\begin{eqnarray}}
\newcommand{\eeqs}{\end{eqnarray}}

\catcode`@=11
\@addtoreset{equation}{section}
\@addtoreset{equation}{subsection}
\def\theequation{\ifnum\value{section}=0 \arabic{equation}\ignorespaces
\else \ifnum\value{section}=-1 A.\arabic{equation}\ignorespaces
\else \ifnum\value{subsection}=0 \thesection.\arabic{equation}\ignorespaces
\else \thesection.\arabic{subsection}.\arabic{equation}\ignorespaces
                           \fi
                      \fi
                 \fi}
\catcode`@=12

\begin{document}

\def\thefootnote{\fnsymbol{footnote}}
\baselineskip 6.0mm

\begin{flushright}
\begin{tabular}{l}
ITP-SB-96-37 \\
August, 1996 \\
\end{tabular}
\end{flushright}

\vspace{8mm}
\begin{center}

{\Large \bf Ground State Entropy and the $q=3$ Potts }

\vspace{2mm}

{\Large \bf Antiferromagnet on the Honeycomb Lattice}

\vspace{16mm}

\setcounter{footnote}{0}
Robert Shrock\footnote{email: shrock@insti.physics.sunysb.edu}
\setcounter{footnote}{6}
and Shan-Ho Tsai\footnote{email: tsai@insti.physics.sunysb.edu}

\vspace{6mm}

Institute for Theoretical Physics  \\
State University of New York       \\
Stony Brook, N. Y. 11794-3840  \\

\vspace{16mm}

{\bf Abstract}
\end{center}
    We study the $q$-state Potts antiferromagnet with $q=3$ on the 
honeycomb lattice. Using an analytic argument together with a Monte Carlo
simulation, we conclude that this model is disordered for all $T \ge 0$. 
We also calculate the ground state entropy to be $S_0/k_B = 0.507(10)$ and
discuss this result. 

\vspace{16mm}

\pagestyle{empty}
\newpage

\pagestyle{plain}
\pagenumbering{arabic}
\renewcommand{\thefootnote}{\arabic{footnote}}
\setcounter{footnote}{0}

  The effect of ground state disorder and associated nonzero ground state
entropy $S_0$ has been a subject of longstanding interest.  A physical 
example is ice, for which $S_0 = 0.82 \pm 0.05$ cal/(K-mole), i.e., 
$S_0/k_B = 0.41 \pm 0.03$ \cite{ice,liebwu}\footnote{Henceforth, we shall use
units such that $k_B \equiv 1$.}. Among spin models, an example is 
the Ising antiferromagnet (AF) on the triangular lattice. In the context of 
this model, Wannier argued that a nonzero ground state (g.s.) entropy 
implies the absence of long-range order, viz., staggered magnetization $M_{st}$
for $T \ge 0$ \cite{wannier}. Another example is the Ising AF on the kagom\'e 
lattice \cite{kn,suto}. In both of these Ising models, the nonzero g.s. 
entropy has the effect of removing a phase transition at
finite temperature.  The Ising AF on the triangular lattice is critical at 
$T=0$ \cite{stephenson}, while on the kagom\'e lattice, with a larger 
value of $S_0$, it is disordered even at $T=0$ \cite{suto}. In these two 
cases, the nonzero g.s. entropy is associated with frustration. However, there
are also spin models, such as the antiferromagnetic $q$-state Potts model 
\cite{potts}-\cite{baxterbook} on the square (sq) and honeycomb (hc) 
lattice, which exhibit g.s. entropy without frustration.  Because of the
absence of frustration, these models constitute ideally simple cases where 
one can study the effects of ground state entropy on the thermodynamics
of a statistical mechanical model.\footnote{Ground state entropy without 
frustration can also occur in models with continuous variables and 
interactions \cite{anem}. A yet more complicated case is that of quenched 
disorder with frustration, as in spin glasses.} 
In contrast to the ferromagnetic (FM) Potts model, which has a 
finite-temperature phase transition for dimensionality $d > 1$, the question of
whether the $q$-state Potts AF has a phase transition at finite (or zero) 
temperature is more delicate and depends on both the value of $q$ and the 
type of lattice. The $q=3$ Potts AF on the square lattice has been well 
studied; an exact 
result of Baxter showed that it is critical at $T=0$ \cite{baxter82}, in 
agreement with a renormalization group argument \cite{ns}, and several Monte
Carlo simulations have been performed on it \cite{wsk,sokal}.  However, to our
knowledge, the behavior of the $q=3$ Potts AF on the honeycomb lattice 
has not been definitely established. We report here the results of a 
study of this model.

  The (isotropic, nearest-neighbor, zero-field) $q$-state Potts model on a
lattice $\Lambda$ is defined by the partition function 
$Z = \sum_{ \{ \sigma_n \} } e^{-\beta {\cal H}}$ with the Hamiltonian
\beq
{\cal H} = -J \sum_{\langle nn' \rangle} \delta_{\sigma_n \sigma_{n'}} 
\label{ham}
\eeq
where $\sigma_n=1,...,q$ are $Z_q$-valued variables on each site $n \in
\Lambda$, $\beta = T^{-1}$, and $J < 0$ for the AF case. 
We define $K = \beta J$, $a=e^K$, $x=(a-1)/\sqrt{q}$, and the reduced free 
energy (per site) $f = -\beta F = \lim_{N \to \infty} N^{-1} \ln Z$, where
$N$ denotes the number of sites in the lattice.  We consider $\Lambda=hc$ here.

   We first observe that for the $q=2$ (Ising) case, the
paramagnetic-ferromagnetic (PM-FM) and PM-AFM critical points are both
determined by the equation \cite{kj} 
\beq
\sqrt{q}+3x-x^3=0 
\label{criteq}
\eeq
These are $a_c=2+\sqrt{3}$ (PM-FM) and $a_{c,AF}=a_c^{-1}=2-\sqrt{3}$
(PM-AFM). (The third root of eq. (\ref{criteq}) is a complex-temperature 
singular point at $a=-1 \equiv a_s$.) 
An equivalent representation of the partition function is 
$Z = \sum_{G' \subseteq G}v^{b(G')}q^{n(G')}$ \cite{kf,baxter73,baxterbook}, 
where $G'$ denotes a subgraph of $G = \Lambda$, $v=(a-1)$, $b(G')$ is the
number of bonds and $n(G')$ the number of connected components of $G'$.  This 
enables one to analytically continue the model from positive integral $q$ to 
real $q$ \cite{baxter73,baxterbook}.  Carrying out this analytic continuation 
and analyzing eq. (\ref{criteq}) for the critical points, one sees that the
points $a_c(q)$ and $a_{c,AF}(q)$ increase and decrease, respectively, 
reflecting the fact that as $q$ increases, one must go to lower temperature 
to achieve FM and AFM long-range order.  As $q$ reaches the value 
$q_z = (3+\sqrt{5})/2 = 2.618..$, \ 
$a_{c,AF}$ decreases to 0, i.e., the AFM phase is squeezed out, and there is 
no longer any finite-temperature AF critical point, which now occurs only at 
$T=0$.  Note that $q_z = B_5 = 1+\tau$, where $B_r=4\cos^2(\pi/r)$ is the
$r'th$ Beraha number \cite{beraha} and $\tau$ is the golden mean. 
For $q > q_z$, $a_{c,AF}$ is negative, i.e., an unphysical, 
complex-temperature (CT) singular point.  It follows that for $q > q_z$ and, in
particular, for $q=3$, the hc Potts AF has no critical point or associated 
continuous phase transition at any $T \ge 0$.\footnote{From these exact
results, recalling the connection $Z(q,\Lambda,K=-\infty)=P_\Lambda(q)$, 
where $P_G(q)$ is the chromatic polynomial for the graph $G$ \cite{chrom}, we
would expect that the behavior of $P_{hc}(q)$ (where $\Lambda = hc$ denotes the
thermodynamic limit of the hc lattice) would differ for $q < q_z$ and $q >
q_z$, as would follow if the zeros of $P_{hc}(q)$ formed a boundary curve in
the complex $q$ plane which crossed the real axis at $q_z$ and separated the
regions which include the segments $q < q_z$ and $q > q_z$. Given the
observation that the crossing point of a boundary curve increases by about 
$\Delta q \simeq 0.4$ from an $8 \times 8$ triangular lattice with cylindrical
boundary conditions (CBC's) to the thermodynamic limit \cite{baxter86}, our 
expectation is consistent with the finding \cite{baxter86} that there is a 
crossing curve on the $8 \times 8$ hc lattice with CBC's at $q \simeq 2.2$.}
As $q$ increases from $q_z$ to $q=3$, the root of eq. (\ref{criteq}) 
which, for $q < q_z$, was the PM-AFM critical point $a_{c,AF}(q)$, moves 
leftward from the origin.  Since for $q > q_z$, there is no longer any physical
AFM phase, we shall denote this point as $a_{c2}$; it moves from
$a_{c2}(q_z)=0$ leftward to $a_{c2}(3)=-0.1848..$.  Meanwhile, as
$q$ increases from 2 to 3, (i) the PM-FM critical point $a_c(q)$ moves to the
right, through $a_c(q_z)=(1/2)(3+\sqrt{15+6\sqrt{5}})=4.1654..$ to 
$a_c(3)=4.4115..$, and the disordered, paramagnetic (PM) phase (and its
complex-temperature generalization) expands accordingly; (ii) the root 
$a_s(q)$ of (\ref{criteq}) moves leftward, from $-1$, through
$a_s(q_z)=(1/2)(3-\sqrt{15+6\sqrt{5}})=-1.1654..$, to $a_s(3)=-1.2267..$ \ .

  In particular, this argument by analytic continuation in $q$ excludes the 
possibility, for the hc lattice, of a massless low-temperature 
phase with algebraic asymptotic decay of correlation functions of the type
discussed in Ref. \cite{bk}. However, this leaves open the possibility that 
the model might have a first-order transition (with finite correlation
length, and hence noncritical). Indeed, this is what did happen for the $q=3$ 
Potts AF on the triangular lattice \cite{grest,entingwu}, although for that
case, the ground state is only finitely degenerate, so that $S_0=0$, and 
$M_{st}$ is nonzero below the transition.  In contrast, given that $S_0$ is 
nonzero in the present case of the hc lattice, the Wannier argument implies 
that $M_{st}=0$ for all $T \ge 0$, so that the discontinuity at such a 
hypothetical transition would have to occur in $U$ but not in $M_{st}$. 
We consider this to render such a transition unlikely but do not know of a 
proof which precludes a discontinuity in short-range order (which enters into
$U$) while long range order, $M_{st}$, remains zero. 

  An effective way to study this possibility of a phase transition is to
perform Monte Carlo simulations of the model, and we have done
this.\footnote{Two other methods would be (i) to analyze high-temperature 
series expansions (a first-order transition could manifest itself in peculiar
behavior of exponents, as in \cite{entingwu}); and (ii) to calculate 
complex-temperature zeros of the partition function and search for a new 
phase boundary which crosses the positive real $a$ axis at a point not 
included in the roots of (\ref{criteq}).  (This relies on the fact that in the
thermodynamic limit, these zeros typically merge to form curves which separate
complex-temperature generalizations of phases. If such a boundary were found,
the density of zeros near the real axis would yield the critical exponent 
$\alpha$, and $\alpha=1$ would suggest a first-order transition.) 
While these methods are of interest in their own 
right, we were able to obtain convincing evidence for our conclusion from the 
Monte Carlo method alone.} For the Monte Carlo simulation, we have used two
different algorithms to update the spins: the Metropolis algorithm and the 
Swendsen-Wang cluster algorithm (SWCA) \cite{sw}.  
Since the SWCA reduces critical slowing down in simulations of
models exhibiting a critical points with divergent correlation lengths, 
the agreement of the results obtained from these two algorithms serves as a 
confirmation of our conclusion from the analytic argument above that the 
model is not critical at $T=0$. For Metropolis, we used lattices with 
periodic boundary conditions (BC's) of sizes ranging from $8 \times 8$ to 
$40 \times 40$. For SWCA, following Ref. \cite{sw}, we used lattices with 
helical BC's in the horizontal direction and free BC's in the vertical 
direction (where the hc lattice is represented as a brick lattice with
horizontal bricks) with sizes ranging from $19 \times 19$ to $39 \times 39$. 
Typically, we ran for several thousand sweeps through the lattice for 
thermalization before calculating averages.  
Each average was calculated from 10,000 sweeps through the
lattice. The full data was obtained as a thermal loop, to test for any
hysteresis associated with either critical slowing down or metastability.  No
such hysteresis was observed. The results obtained with these
two different algorithms were in excellent agreement, differing at most by only
about 1 \%.  In Fig. 1 we show measurements of the internal energy per site,
$U$, from the SWCA simulation.  

\begin{figure}
\epsfxsize=3.5in
\epsffile{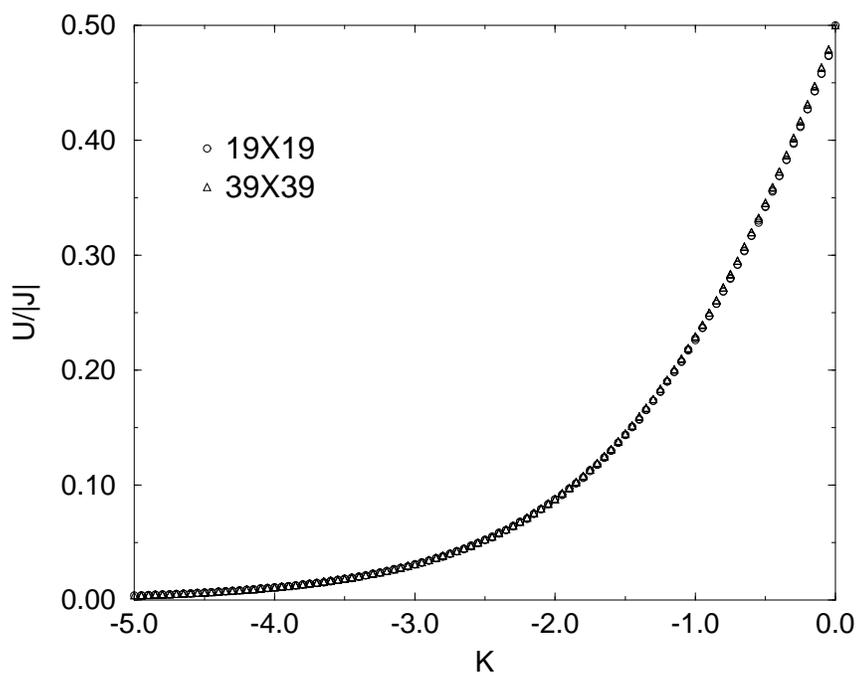}
\caption{Measurements of internal energy $U$, as a function of
$K=\beta J$, for the $q=3$ Potts antiferromagnet on the honeycomb 
lattice. See text for details.}
\label{fig1}
\end{figure}
The intercept and slope at $K=0$ follow from the high-temperature expansion 
$-U/J = (g/2)\Bigl [ 1/q + ((q-1)/q^2)K + O(K^2) \Bigr ]$, where 
$g$ is the coordination number of the lattice, so 
$U/|J|=1/2 + (1/3)K + O(K^2)$ for the $q=3$ Potts AF on the hc lattice. 
Clearly, at $T=0$,
i.e., $K=-\infty$, the spins on each bond must be different, so 
$U=-(g/2)\langle \delta_{\sigma_n \sigma{n'}}\rangle = 0$.  As a check on
our program, we have also simulated the $J > 0$ model and obtained excellent
agreement with the PM-FM phase transition known from eq. (\ref{criteq}) 
to occur at $K_c=1.484$. 
Evidently, the data smoothly curves down from the $K=0$ value toward the 
$K=-\infty$ value as $K$ decreases to $-5$; there is no indication of any 
phase transition, in particular, a first-order one.  The absence of
any critical slowing-down for large negative $K$ is in agreement with our
analytic argument from eq. (\ref{criteq}) that the model is not critical at
$T=0$.  The fact that the data for the $19 \times 19$ and $39 \times 39$ 
lattices are very close to each other (as was also true for the intermediate 
sizes that we used) shows that it is not
necessary to go to larger lattice sizes; the present ones are adequate
for our conclusion.  Indeed, this is not surprising, in view of our result 
that the lattice is disordered for $T \ge 0$ (if there had been any indication
of critical behavior as signalled, e.g. by critical slowing-down, then we would
also have run simulations on larger lattices).

     Our results imply that the $q=3$ Potts AF on the hc lattice has the 
property that in the complex $a$ plane, the points $a=1$ ($K=0$) and $a=0$ 
($K=-\infty$) are analytically connected and $a=0$ does not lie on a
complex-temperature (CT) phase boundary. We have calculated CT zeros of $Z$ 
on small lattices with periodic boundary conditions and have obtained results
which are consistent with this conclusion.\footnote{Calculations of 
CT zeros for this model are also being performed by A. J. Guttmann and I. 
Jensen. We thank these authors for informing us of their work.} We 
note that previous studies of the CT zeros of the square-lattice Potts model 
for $q=3$ and 4 have shown that the pattern of zeros in the $Re(a) < 0$ 
region exhibits a significant dependence on the boundary conditions 
\cite{martin}-\cite{p}. 

    For $q \ge 4$, it has been proved that the Potts AF on the hc lattice is 
disordered for all $T \ge 0$ \cite{sokalbound}.  This result is quite 
consistent with our finding, since increasing $q$ increases the disorder in 
the model. 

   As part of our study, we have calculated the g.s. entropy
$S_0(\Lambda,q)=S_0(hc,3)$ of the model, using the relation 
\beq
S(\beta) = S(\beta=0) + \beta U(\beta) - \int_0^{\beta} U(\beta')d\beta'
\label{seq}
\eeq
starting the integration at $\beta=0$ with $S(\beta=0)=\ln q$ for the 
$q$-state Potts model.  We found, as in previous work \cite{binder}, that 
this provides a very accurate method for calculating $S_0$. For this we used 
the Metropolis algorithm with periodic BC's for several lattice sizes.  
Since $U(K)$ rapidly approaches its asymptotic value of 0 as $K$ decreases 
past about $K=-5$ (see Fig. 1), the RHS of (\ref{seq}) rapidly approaches a
constant in this region, enabling one to obtain the resultant value of 
$S(\beta=\infty)$ for each lattice size.  We then performed a fit to this data 
and extrapolated the result to the thermodynamic limit; the results are shown
in Fig. 2. 
\begin{figure}
\epsfxsize=3.5in
\epsffile{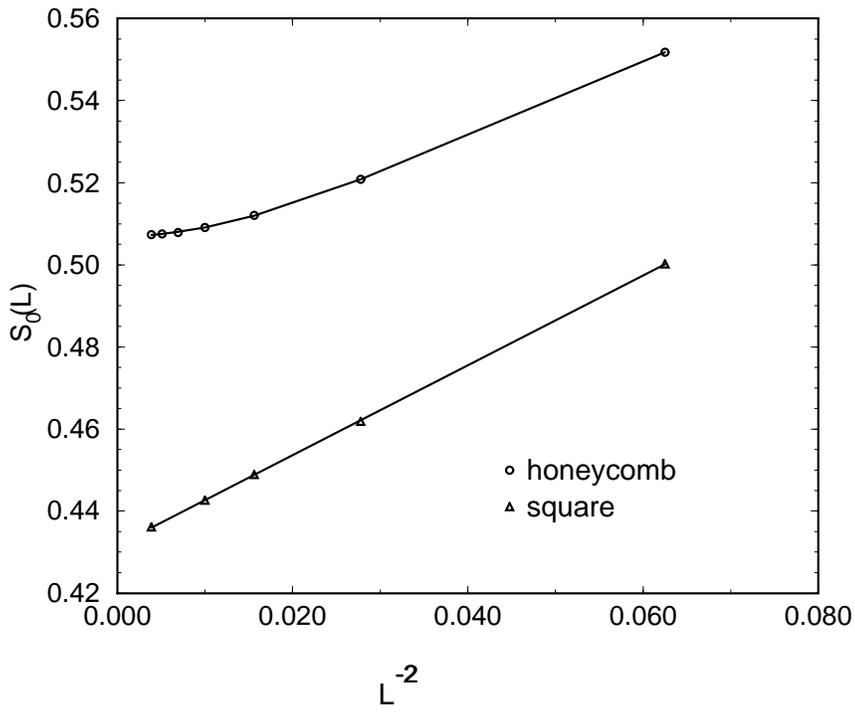}
\caption{Measurements of ground state entropy $S_0$, as a function of lattice
size, for the $q=3$ Potts AF on the honeycomb and square lattices.} 
\label{fig2}
\end{figure}
As a check, we also carried out the analogous calculations for the $q=3$
Potts AF on the square lattice.  A fit to the finite-size dependence 
of our sq lattice data agrees very well with the form found in 
Refs. \cite{wsk} and \cite{pw}, 
$S_0(sq,3) = S_0(L;sq,3) + c_{sq,3}/L^2$ with 
$c_{sq,3}=1.077$, and we get 
$S_0(sq,3)=0.4317(3)$, in excellent agreement with the exact value 
$S_0(sq,3) = (3/2)\ln(4/3)=0.4315...$ \cite{liebwu,wurev,baxterbook}. 
For the hc lattice, as is evident from Fig. 2, our measurements do not 
exhibit the same finite-size dependence as for the sq lattice. An 
empirical function including terms up to $L^{-6}$ yields a good fit to the 
data (see Fig. 2) and gives the $L=\infty$ value of the g.s. entropy for the 
$q=3$ hc Potts AF 
\beq
S_0(hc,3) = 0.507 \pm 0.010
\label{s0}
\eeq
where the error is an estimate of the uncertainty. This yields
$W(hc,3)=1.66 \pm 0.02$, where $W(\Lambda,q) = e^{S_0(\Lambda,q)}$.  
We observe that our results are consistent, to within the uncertainty, with 
the exact expression $W(hc,3)=5/3$.  The ratio 
$R_S(\Lambda,q) = S(\Lambda,q,T=0)/S(q,T=\infty)$ serves as a useful 
measure of the reduction of disorder in a given model as $T$ decreases from 
$\infty$ to 0.  Our results yield $R_S(hc,3)=0.4615 \pm 0.010$ for the $q=3$
hc Potts AF, which shows that the disorder at $T=0$ is a substantial fraction 
of its maximal, $T=\infty$ value. 

  From the basic relation $S = \beta U + f$ and the property that
$\lim_{K \to -\infty}\beta U(\beta) = 0$, as is true of the $q$-state Potts AF
models considered here, it follows that 
\beq
S_0(\Lambda,q) = f(\Lambda,q,K=-\infty) = \lim_{N \to \infty}N^{-1}
\ln(P_\Lambda(q))
\label{sp}
\eeq
or equivalently $W(\Lambda,q) = \lim_{N \to \infty} N^{-1} P_\Lambda(q)$. 
That is, the g.s. entropy is determined by the asymptotic behavior of the 
chromatic polynomial in the thermodynamic limit. Series of the form 
$W(\Lambda,q) = q\Bigl ( (q-1)/q \Bigr )^{g/2}\bar W(\Lambda,q)$, where 
$\bar W(\Lambda,q)=1+\sum_{n=1}^\infty w_n y^n$ with $y=1/(q-1)$ were 
calculated in Ref. \cite{kewser}. 
It is of interest to compare our result (\ref{s0}) with an 
estimate from the series for $\bar W(hc,3) = 1 + y^5 + 2y^{11} + 4y^{12} 
+ ...$, calculated through $O(y^{18})$ \cite{kewser}.  Because of the sign
changes in the hc series (the coefficients of the first five terms are 
positive, while those of the remaining four terms are negative), it is 
difficult to make a reliable extrapolation.  Simply taking the sum yields 
$W(hc,3)=1.687$, which is agreeably close to our central value, 1.66. 

   In summary, combining analytic arguments and a Monte Carlo simulation, we
have reached the conclusion that the $q=3$ Potts AF on the honeycomb lattice is
disordered for all $T \ge 0$ and have calculated the ground state entropy for
this model. 

This research was supported in part by the NSF grant PHY-93-09888.

\vfill
\eject

\vspace{6mm}

\begin{thebibliography}{99}

\bibitem{ice}{Giauque, W. F. and Stout, J. W. 1936, J. Am. Chem. Soc. 
{\bf 58}, 1144; Pauling, L. 1960, {\it The Nature of the Chemical Bond}
(Cornell Univ. Press, Ithaca), p. 466.} 

\bibitem{liebwu}{Lieb, E. H. and Wu, F. Y. 1972, in C. Domb and M. S. Green, 
eds., {\it Phase Transitions and Critical Phenomena} (Academic Press, 
New York) v. 1, p. 331.}

\bibitem{wannier}{Wannier, G. H. 1950, Phys. Rev. {\bf 79}, 357.}

\bibitem{kn}{Kano,K. and Naya, S. 1953, Prog. Theor. Phys. {\bf 10}, 158.}

\bibitem{suto}{S\"uto, A. 1981, Z. Phys. B {\bf 44}, 121.} 

\bibitem{stephenson}{Stephenson, J. 1964, J. Math. Phys. {\bf 5}, 1009;
Stephenson, J. 1970 {\it ibid.} {\bf 11}, 420.}

\bibitem{potts}{Potts, R. B. 1952, Proc. Camb. Phil. Soc. {\bf 48}, 106.}

\bibitem{wurev}{Wu, F. Y. 1982, Rev. Mod. Phys. {\bf 54}, 235.}

\bibitem{baxterbook}{Baxter, R. J. 1982, {\it Exactly Solved Models in 
Statistical Mechanics} (Academic Press, New York).}

\bibitem{anem}{Kohring, G. and Shrock, R. 1988, Nucl. Phys. B {\bf 295}, 36.}

\bibitem{baxter82}{Baxter, R. J. 1982, Proc. Roy. Soc. London, Ser. A
{\bf 383}, 43.}

\bibitem{ns}{Nightingale, M. P. and Schick, M. 1982, J. Phys. A, {\bf 15}, L39;
den Nijs, M. P. M., Nightingale, M. P., and Schick, M. 1982, Phys. Rev. B 
{\bf 26}, 2490.}

\bibitem{wsk}{Wang, J.-S., Swendsen, R. H., and Koteck\'y, R. 1989, Phys. 
Rev. Lett. {\bf 63}, 109; 1990, Phys. Rev. B {\bf 42}, 2465.}

\bibitem{sokal}{Ferreira, S. J. and Sokal, A. D. 1995, Phys. Rev. B {\bf 51}, 
6727.} 

\bibitem{kj}{Kim, D. and Joseph, R. J. 1974, J. Phys. C {\bf 7}, L167; 
Baxter, R. J., Temperley, H. N. V., and Ashley, S. 1978, Proc. Roy. Soc. 
London, Ser. A {\bf 358}, 535; Burkhardt, T. W. and Southern, B. W. 1978, 
J. Phys. A {\bf 11}, L247.} 

\bibitem{kf}{Kasteleyn, P. W. and Fortuin, C. M. 1969, J. Phys. Soc. Jpn. 
Suppl. {\bf 26}, 11; Fortuin, C. M. and Kasteleyn, P. W. 1972, Physica 
{\bf 57}, 536.} 

\bibitem{baxter73}{Baxter, R. J. 1973, J. Phys. C {\bf 6}, L445.}

\bibitem{beraha}{Beraha, S., Kahane, J. and Weiss, N. 1980, J. Combin. 
Theory B {\bf 28}, 52.} 

\bibitem{chrom}{Read, R. C. 1968, J. Combin. Theory {\bf 4}, 52; 
Read, R. C. and Tutte, W. T. 1988, in Beineke, L. and Wilson, R., eds. 
{\it Selected Topics in Graph Theory} (Academic Press, New York), 
v. 3, p. 15.}

\bibitem{baxter86}{Baxter, R. J. 1986, J. Phys. A {\bf 20}, 5241.}

\bibitem{bk}{Berker, A. N. and Kadanoff, L. P. 1980, J. Phys. A {\bf 13}, 
L259.} 

\bibitem{grest}{Grest, G. S. 1981, J. Phys. A {\bf 14}, L217 (we thank
G. Grest for a discussion); Saito, Y. 1981, J. Phys. A {\bf 15}, 1885; 
Adler, J., Brandt, A., Janke, W., and Shmulyian, S. 1995, J. Phys. A 
{\bf 28}, 5117.} 

\bibitem{entingwu}{Enting, I. G. and Wu, F. Y. 1982, J. Stat. Phys. 
{\bf 28}, 351.} 

\bibitem{sw}{Swendsen, R. H. and Wang, J.-S. 1987, Phys. Rev. Lett. 
{\bf 58}, 86; Wang, J.-S. and Swendsen, R. H.  1990, Physica A {\bf 167}, 
565.} 

\bibitem{martin}{Martin, P. P. 1991, {\it Potts Models and Related Problems in
Statistical Mechanics} (World Scientific, Singapore).}

\bibitem{wuzero}{Chen, C. N., Hu, C. K., and Wu, F. Y. 1996, Phys. Rev. Lett. 
{\bf 76}, 169.}

\bibitem{p}{Matveev, V. and Shrock, R., Phys. Rev. in press 
(cond-mat/9605176).}

\bibitem{sokalbound}{Salas, J. and Sokal, A. D., NYU preprint 
cond-mat/9603068.} 

\bibitem{binder}{Binder, K. 1981, Zeit. f. Physik B {\bf 45}, 61.}

\bibitem{pw}{Park, H. and Widom, M. 1989, Phys. Rev. Lett. {\bf 63}, 1193.} 

\bibitem{kewser}{Kim, D. and Enting, I. G. 1979, J. Combin. Theory, B {\bf 26},
327.} 

\end{thebibliography}
\end{document}